\documentclass[twocolumn,showpacs,preprintnumbers,amsmath,amssymb,eps]{revtex4}
                           
\usepackage{graphicx}
\usepackage{dcolumn}
\usepackage{bm}

\begin{document}

\preprint{APS/123-QED}

\title{Upper critical field $H_{c2}$ calculations for the high critical temperature
superconductors considering inhomogeneities}

\author{E. S. Caixeiro}
\author{J. L. Gonz\'alez}
\author{E. V. L. de Mello}
\affiliation{%
Departamento de F\'{\i}sica, Universidade Federal Fluminense, Niter\'oi, RJ 24210-340, Brazil\\}%


\date{\today}

\begin{abstract}

We perform calculations to obtain the $H_{c2}$ curve
of high temperature superconductors (HTSC). We consider 
explicitly the fact that the HTSC possess intrinsic inhomogeneities
by taking into account a non uniform charge density $\rho(r)$. 
The transition to a coherent superconducting phase at a critical temperature $T_c$ corresponds
to a percolation threshold among different superconducting regions, each one characterized by
a given  $T_c(\rho(r))$. Within this model we calculate the upper 
critical field $H_{c2}$ by means of an average linearized Ginzburg-Landau (GL) equation 
modified to take into account the distribution of local superconducting
temperatures $T_c(\rho(r))$. This approach explains some of the anomalies associated 
with $H_{c2}$ and why several properties like the Meissner and
Nernst effects are detected at temperatures much higher than $T_c$.

\end{abstract}

\maketitle

\section{Introduction}
High critical temperature superconductors (HTSC) have been discovered fifteen years ago~\cite{Bednorz}, 
but many of their properties remains not well explained. Some of their features 
are completely different from the low temperature superconductors, for example,
the $H$-$T$ phase diagram of the HTSC possess, in certain cases, positive
curvature for $H_{c2}(T)$, with no evidence of saturation at low temperatures
~\cite{Positivecurve4, Positivecurve2, Positivecurve1}. 
These lack of saturation at low temperatures may minimize the importance of 
strong fluctuations of the order parameter which was one of the earlier leading 
ideas~\cite{Positivecurve3,Blatter}. Furthermore, this behavior cannot be explained 
by the classical WHH (Werthamer-Helfand-Hohenberg)
theory for superconductors~\cite{WHH1, WHH2} where
$H_{c2}(T)$ exhibits negative curvature and saturates at low temperatures. 

In order to interpret this unusual behavior and furnishes ground to the $H_{c2}$ positive curvature 
many different possibilities were proposed like, for instance: the interplay between Anderson localization 
and superconductivity which affects the coherent length near the mobility edge~\cite{Kotliar}, the 
effect of correlations among magnetic impurities which leads to a weakening of the pair breaking 
process~\cite{Ovchinnikov}, mesoscopic fluctuations~\cite{Spivak}, depletion of vortex viscosity 
or local pairs confined to small superconducting regions coupled through a normal host due to 
anisotropic quasiparticle relaxation rate around the Fermi surface~\cite{Geshkenbein}, 
effective two component (boson-fermion) model~\cite{Domanski}, Luttinger liquid behavior in the normal
state~\cite{Schofield, Wheatley}, bipolaron superconductivity~\cite{Alexandrov}, 
charge-density waves and stripes~\cite{Maska}.

On the other hand, 
another feature that has recently received considerable attention is the intrinsic inhomogeneous 
distribution of charge (or doping level) and superconducting gaps, as demonstrated by the STM/S 
experiments~\cite{Pan,Lang,Howald}. Based on these 
measurements and in a number of others, like neutron 
diffraction~\cite{Egami,Egami2,Billinge,Bozin}, we have proposed a 
general theory to explain the main features of the 
HTSC phase diagram~\cite{Evandro1,Evandro2}. The main point is that near the pseudogap temperature 
$T^*$~\cite{Timusk} some localized regions become superconducting, and the size of these regions 
increases as the temperature is lowered and, at the critical temperature $T_c$, these 
superconducting regions percolate,and the system becomes able to hold a dissipationless current.
This scenario is consistent with the Meissner effect above $T_c$ 
measured by Iguchi et al.~\cite{Iguchi} and the diamagnetic signal above $T_c$ which has also been 
measured by other groups by susceptibility~\cite{Lascialfari}. The drift of magnetic vortices 
expected to occur in the a superconductor phase has also been measured in HTSC at temperatures 
above $T_c$ through the Nernst effect~\cite{Yayu1}.

In this paper we develop a unified view for all these anomalous properties
to explain the observed features of the upper critical field 
$H_{c2}$ unusual properties in HTSC. In our approach the pseudogap temperature $T^*$ is where small static
superconducting regions starts to develop~\cite{Evandro1}. The intrinsic inhomogeneities in the charge 
produces superconducting gaps which varies locally inside a given sample and, as the temperature 
decreases below $T^*$, the superconducting regions start to grow as it has been demonstrated by the 
Meissner effect measurements~\cite{Iguchi}. 
At temperatures below $T^*$, but above the critical temperature $T_c$, there is no long range order 
but there are several isolated local superconducting regions inside the sample. This scenario 
explains why there is some indications that $H_{c2}$ does not vanish at $T_c$ ~\cite{Yayu} but at a 
much larger value; also a much larger magnetization than the expected, above $T_c$, 
has been recently measured~\cite{Lascialfari}. 
The main difficulty to perform calculations in this scenario is the lack of information on the 
inhomogeneous distribution of charge in a given sample. Based on the experimental results which led 
to the idea of stripes~\cite{Trancada,Bianconi} and on the STM/S results, we have proposed a 
bimodal distribution for the charge inside a given HTSC~\cite{Evandro1}. 
Although we do not know the exact form of such distribution, it mimics the 
antiferromagnetic (AF) insulator regions (insulator branch) and the metallic regions (metallic branch) 
and it contains several features found in the cuprates. Here we will apply such
distribution derived in Ref.~\cite{Evandro1} to calculate the contribution of resulting 
superconducting regions to estimate the magnetic response
for some materials and compare with the experimental results.
This is accomplished through a simple model to compute the contribution of the
superconducting regions to the upper critical field $H_{c2}$, considering that each of these regions acts 
as independent superconducting regions whose $H_{c2}$ is described by the GL equation. 
All the contributions from different regions are computed to give the total upper critical field of a given 
superconducting compound.
This is a clear phenomenological approach that, according to the GL theory, yield good results near the 
critical temperatures. 

It is worthwhile to mention that Ref.~\cite{Geshkenbein} has also inferred that 
the positive curvature of $H_{c2}$ could be due to pair formation in small grains with local $T_c$ higher 
than the bulk $T_c$. Therefore, we can see that several attempts had been 
published~\cite{Ovchinnikov,Spivak,Geshkenbein,Domanski} connecting $H_{c2}$ and its features with the 
effects of some type of disorder. By the same token, we apply below our recently developed 
theory~\cite{Evandro1} on the inhomogeneities of cuprates and pseudogap phenomenon 
to evaluate a theory for the $H_{c2}$ of these materials.

This paper is divided as follows: In Sec.II we present the density of charge distribution 
and the phase diagram of a selected compound of the LSCO family. In Sec.III the upper critical 
field $H_{c2}$ from the GL theory is considered and generalized with the 
inclusion of the inhomogeneous superconducting regions. In Sec.IV we compare the 
theoretical results with some selected experimental data of the LSCO family and
optimum Bi2212 high-$T_c$. A good qualitative agreement is observed. In Sec.V we make the final analyses and 
conclusions.

\section{The density of charge distribution}

Just for completeness, we will briefly outline the basic ideas concerned with the inhomogeneous
charge distribution introduced in Ref.~\cite{Evandro1}.

To model this inhomogeneous medium we consider a phenomenological 
distribution of probability $P(\rho(r))$ of a
given local charge
density $\rho(r)$. The differences in the local charge densities 
yield insulator and metallic regions. For simplicity we hereafter make $\rho(r)$=$\rho$.

The distribution $P(\rho(r))$ we consider is a combination of a Gaussian and a Poisson distribution, 
which becomes the appropriate distribution to deal with the high and low density 
compounds~\cite{Evandro1}, that is, the whole phase diagram.
The main features of $P(\rho)$ is that it has two branches: an insulating one with $0\le\rho\le0.05$, 
and a metallic one which starts at $\rho_m$. 

\begin{figure}[h]
\begin{center}
\includegraphics[height=7cm]{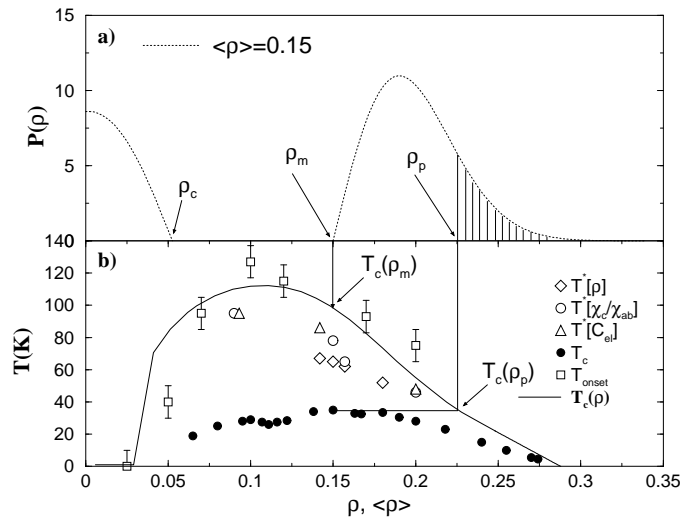}
\caption{a) The distribution $P(\rho)$ for the LSCO $\langle \rho \rangle$=0.15 
compound is shown. The arrows indicates the density $\rho_c$=0.05, 
the percolation density $\rho_p$=0.225, and $\rho_m$, with $\rho_m$=$\langle \rho \rangle$. 
The hachured part indicates the region below the percolation threshold $\rho_p$, i.e. the region below 
the superconducting critical temperature $T_c$.
b) The theoretical local superconducting critical temperatures $T_c(\rho)$ for this compound
is shown as the solid line, together with the
experimental pseudogap temperatures $T^*$ of Refs.~\cite{Xu,Oda}. 
The filled black circles are the experimental critical temperatures $T_c$ of the 
LSCO family from Ref.~\cite{Oda}. The vertical and horizontal lines 
indicates  $T_c(\rho_p)$=$T_c$.}
\label{FigHc21}
\end{center}
\end{figure}
    
For most compounds $\rho_m$$\approx$$\langle\rho\rangle$, where $\langle\rho\rangle$ is the 
average density of a compound. 
In Fig. \ref{FigHc21} we show the phase diagram for the LSCO $\langle\rho\rangle$=0.15 compound 
together with the charge distribution and experimental data of Ref.~\cite{Xu,Oda}.
Since the local critical temperature $T_c(\rho)$ is a decreasing function 
of $\rho$, the maximum $T_c$ is $T_c(\langle\rho\rangle)$, which is therefore the system pseudogap 
temperature $T^*$. Upon cooling below $T^*$ part of the metallic regions become superconducting. 
At $T$=$T_c$ $60\%$ of the material is in the superconducting phase and we say that the 
superconducting regions percolate~\cite{Evandro1}. 

\begin{figure}[h]
\begin{center}
  \begin{minipage}[b]{.1\textwidth}
    \begin{center}
    \centerline{ \includegraphics[width=6.5cm]{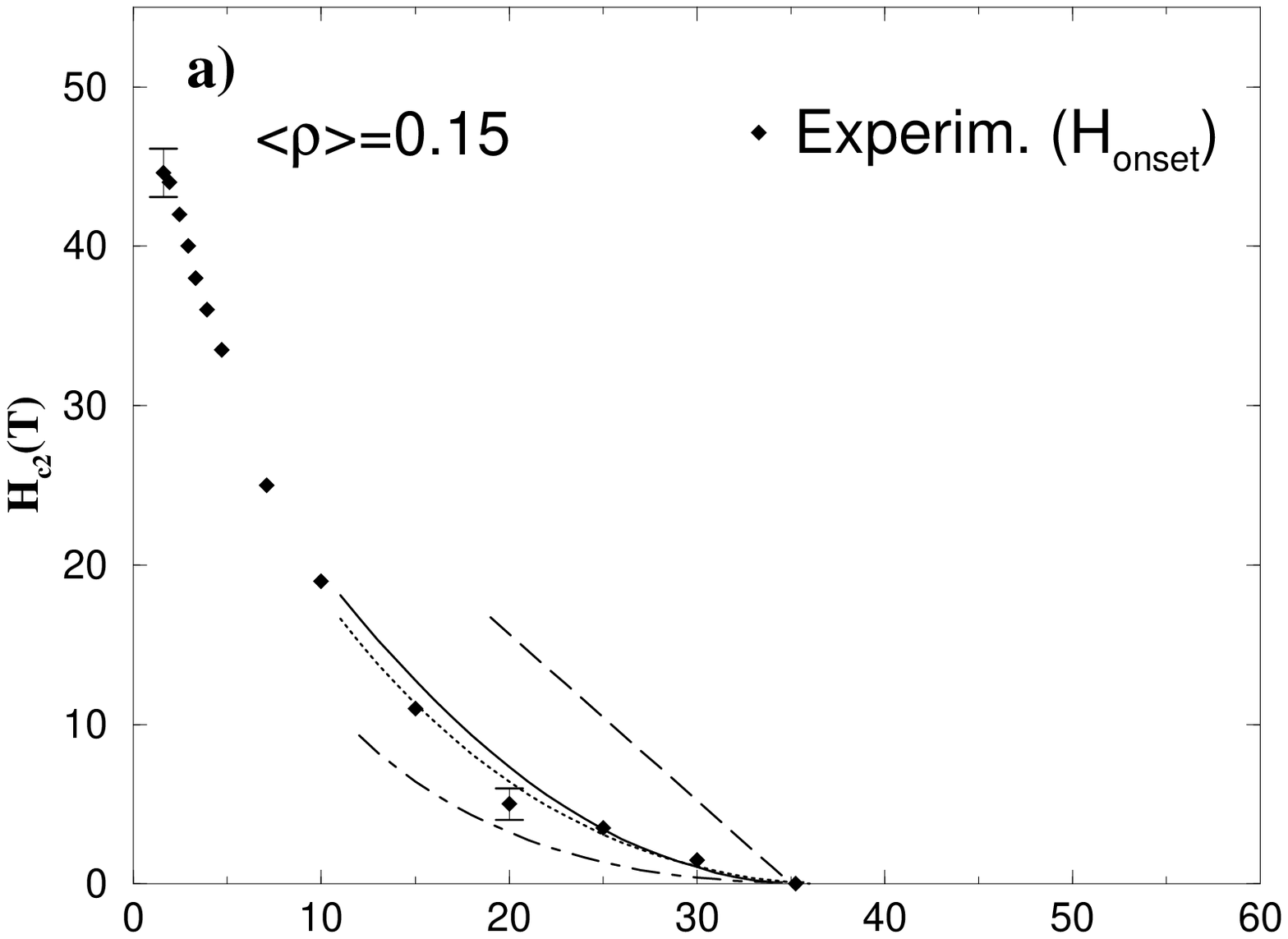}}
    \centerline{ \includegraphics[width=6.5cm]{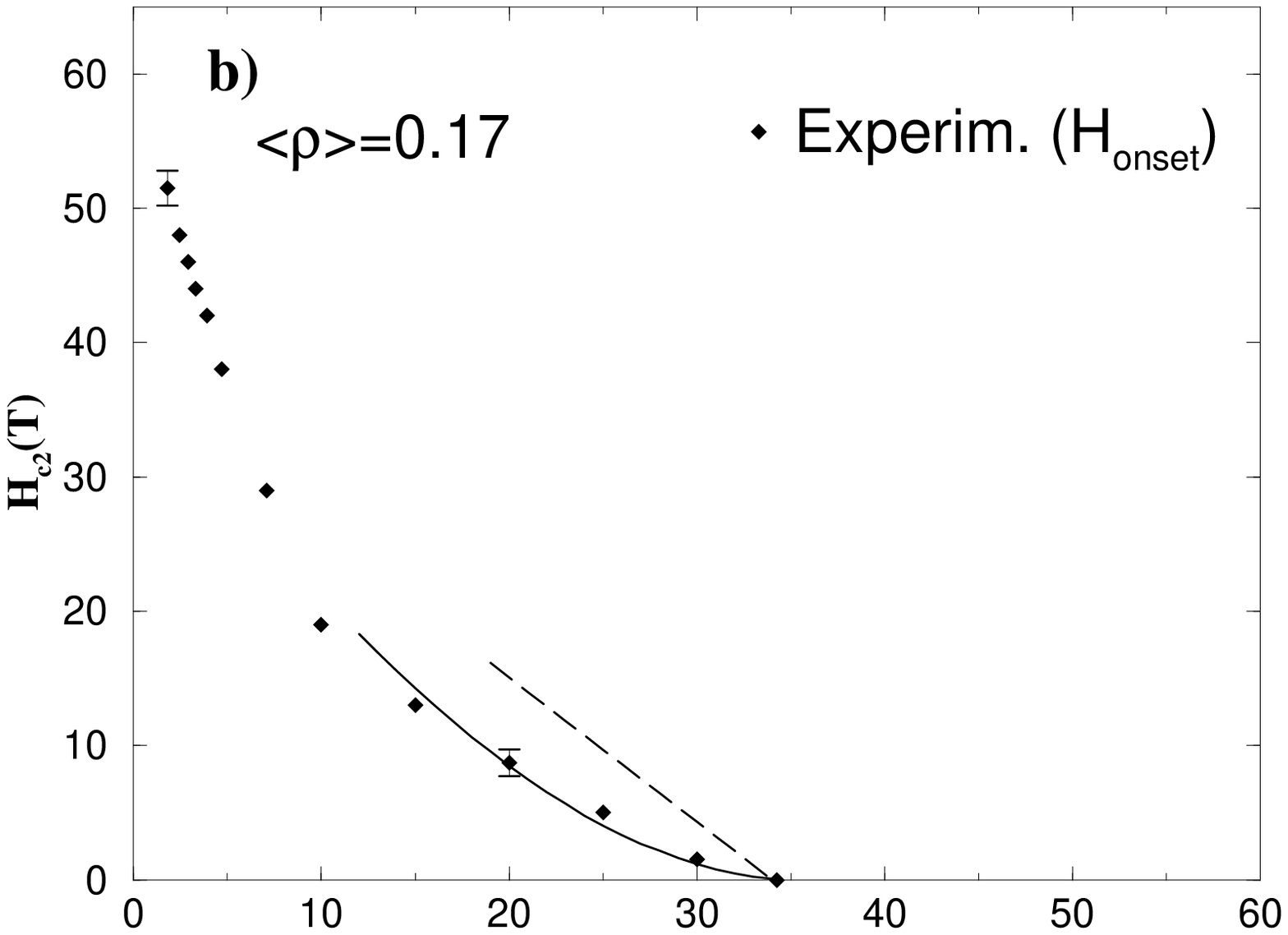}}
    \centerline{ \includegraphics[width=6.5cm]{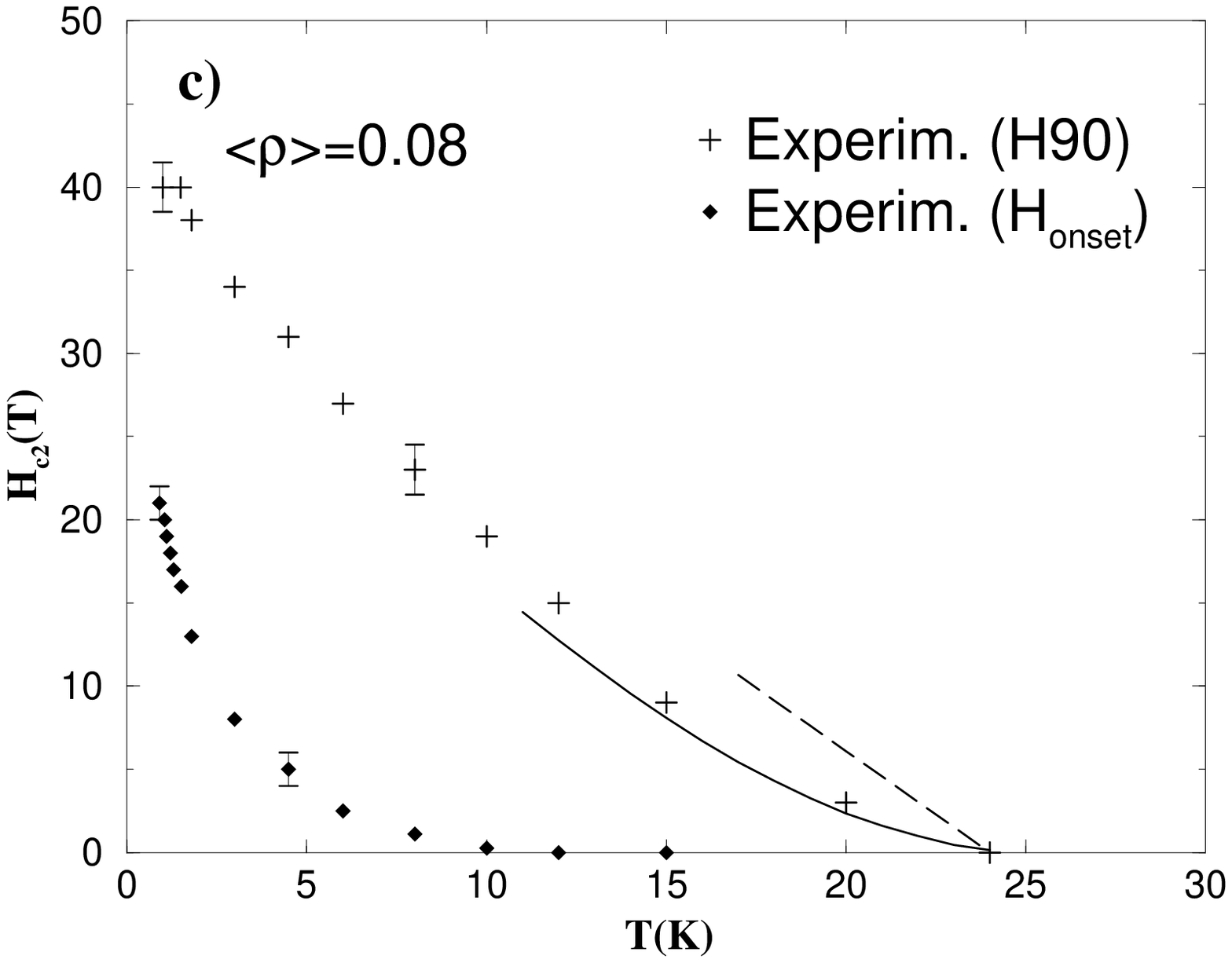}}
    \end{center} 
  \end{minipage}
  \caption{Theoretical results of $H_{c2}(T)$ (solid lines) of the
LSCO series considering the distribution of Ref.~\cite{Evandro1} 
together with the experimental data of Ref.~\cite{Positivecurve4}.
The dashed line is a GL fitting of Eq.(\ref{3}). In a) the results for a constant (dot-dashed line) and 
a linear (dotted line) distribution are also shown.}
\label{Fig2}
 \end{center}
\end{figure}

In order to study the effect of the charge distribution on our results we have considered a constant 
and a linear decreasing distribution, with metallic and insulating branches, evaluated with the densities 
$\rho_c$ and $\rho_p$ (see Fig.\ref{FigHc21}) for $\rho_m$=0.15, and in the same range.  
The results are shown in Fig.\ref{Fig2}a of section IV and indicate that the qualitative behavior remains 
the same, i.e., the inhomogeneities seem to be the cause of the positive curvature, but the quantitative 
agreement is worst than the results from the distribution derived by the STM experiments and used in our 
calculations.  

\section{The Calculations}
It is well known that most of the HTSC are type-II superconductors~\cite{Fisher}. 
For these type of superconductors there are two critical fields in the $H$-$T$ phase diagram: 
the lower $H_{c1}$ and the upper $H_{c2}$.
Above $H_{c2}$ the material returns to the normal metal state.
By definition, one expects the superconductivity to disappear above the upper 
critical field $H_{c2}$. 

In the case of an external magnetic field
parallel to the $c$-direction, i.e. perpendicular to the $CuO_2$ planes ($ab$-direction),
the GL upper critical field is given by~\cite{Pav, Comprimcorr, LD}
\begin{eqnarray}
 H_{c2}(T)&=&{\Phi_{0}\over 2\pi \xi_{ab}^2(T)}
\label{1}
\end{eqnarray}
where $\Phi_0={hc/ 2e}$ is the flux quantum and $\xi_{ab}(T)$ is the GL temperature dependent
coherence length in the $ab$ plane~\cite{Comprimcorr, Blatter, Franz}. Therefore, 
$H_{c2}$ is determined by the coherence length $\xi_{ab}(T)$ of the supercondutor, which is 
treated as a phenomenological parameter. In terms of the GL parameters the coherence length is given
\begin{eqnarray}
\xi_{ab}^2(T)={\hbar^2 \over 2m_{ab}\alpha(T)}=
\xi_{ab}^2(0)\left( {T_c\over T_c-T}\right)\hspace{0.5cm} (T<T_c)
\label{2}
\end{eqnarray}
where $\xi_{ab}^2(0)$=$\hbar^2/2m_{ab}aT_c$ is the extrapolated coherence length, $m_{ab}$ is the 
part of the mass tensor for the $ab$ plane and $a$ is a constant~\cite{Pav}. Using the BCS formula
$\xi \sim v_F/k_BT_c$ we expect a shorter coherence length for HTSC relative to the
low temperature supercondutors due to their 10 times
higher $T_c$'s. However, due to the lower density of carriers~\cite{Pav}, $v_F$ in these materials
is also small, which results in very short coherence length, $\xi \sim 10\AA$. A typical value
for the extrapolated coherence length which we use in our calculations is $\xi_{ab}(0)\sim 15\AA$
and $\xi_{c}(0)\sim 4\AA$ for the YBCO~\cite{Pav}
and $\xi_{ab}(0)\sim 32\AA$ and $\xi_{c}(0)\sim 7\AA$ for LSCO~\cite{Pav, Qiang}, where $\xi_{c}(0)$
is the coherence length in the $c$ direction. One should note that $\xi_{c}(0)$ is smaller than
$\xi_{ab}(0)$ and is of the order of the spacing between adjacent conducting $CuO_2$ planes.

Therefore, the GL upper critical field may be written as
\begin{eqnarray}
H_{c2}(T)&=&{\Phi_{0}\over 2\pi\xi_{ab}^2(0)}\left( {T_c-T\over T_c}\right).\hspace{0.5cm} (T<T_c)
\label{3}
\end{eqnarray}
 
Let us now apply this expression to a HTSC with intrinsic inhomogeneities in the charge 
distribution. 

When considering the inhomogeneity of the HTSC at temperatures below $T^*$, isolated  superconducting 
regions may exist in the form of separated islands
even in magnetic fields $H>H_{c2}(T)$~\cite{Landau1,Landau2}.
For temperatures above $T_c$ there is no long range order but, due to the various different 
local values of $T_c(\rho(r))$ superconducting regions may exist in the form of separated regions. 
Here we calculate the upper critical field $H_{c2}$ for a given sample assuming that each 
isolated or connected superconducting region displays a local $H_{c2}^i$ which is given by the linearized
GL equation with effective mass tensor~\cite{Comprimcorr,Ginzburg}.
Since a given local superconducting region ``$i$'' has a local temperature $T_c(i)$ with a probability 
$P_i$ and local coherence length $\xi_i$, it will contribute to the upper critical field 
with a local linear upper critical field $H_{c2}^i(T)$ near $T_c(i)$. 
Therefore, the total contribution of the local superconducting regions to the upper critical field 
is the sum of all the $H_{c2}^i(T)$'s. Thus, applying Eq.(\ref{3}), the $H_{c2}$ for an 
entire sample is 
\begin{eqnarray}
H_{c2}(T)&=&{\Phi_{0}\over 2\pi\xi_{ab}^2(0)}{1\over W}\sum_{i=1}^N P_i
\left( {T_c(i)-T\over T_c(i)}\right)\nonumber \\
&=&{1\over W}\sum_{i=1}^N P_iH_{c2}^i(T)\hspace{0.25cm} (T<T_c(i)\le T_c)
\label{4}
\end{eqnarray}
where $N$ is the number of superconducting regions, or superconducting 
islands each with its local $T_c(i)$$\le T_c$ and $W=\sum_{i=1}^N P_i$ is the sum 
of all the $P_i$'s. 
As we have already mentioned, at temperatures above $T_c$ there are isolated 
superconducting regions, while below 
$T_c$ these regions percolate and the system may hold a dissipationless current. 
Since $H_{c2}$ is experimentally measured at $T<T_c$($H$=0), it is the field which destroys the 
superconducting clusters with $T<T_c(i)\le T_c$, leading the system without percolation.
The first regions which are broken are the weakest ones, which have critical temperatures $T_c(i)$'s 
lower than $T_c$($H$=0).
The mechanism is the following: 
at a temperature $T<T_c$ most of the system is superconducting and a small applied field destroys first 
the superconducting regions at lower $T_c(i)$'s, without loss of long range order. Increasing the 
applied field   
causes more regions to become normal and eventually when the regions with $T_c(i)\approx T_c$ turn to the 
normal phase, the system is about to have a nonvanishing resistivity. This value of the applied field 
is taken as the $H_{c2}$ in our theory, and it is the physical meaning of Eq.(\ref{4}). Thus, 
at a given temperature $T$, we sum the superconducting regions with $T<T_c(i)\le T_c$, with 
their respective probabilities. 

It is important to notice that the STM experiments have demonstrated that the size of a region with 
constant superconducting gap is of the order of $20\AA$~\cite{Pan,Lang,Howald}. These results have also 
been obtained by $\mu$SR experiments~\cite{Uemura}. According to the discussion in the introduction, at $T^*$ 
some small superconducting isolated islands (or droplets) start to appear through the system. Therefore, 
near $T^*$ the GL approach should not be valid because the size of an island coherence lenght with 
$T_c(i)$$\approx$$T^*$ is of the same order of the superconducting islands. As the temperature decreases the 
superconducting islands start to unite forming larger inhomogeneous superconducting regions. At $T_c$ they 
percolate, occupying 60$\%$ of the system. At $T<T_c$ the occupied superconducting volume is comparable to
the size of the system and clearly much larger than the typical coherence length $\xi_i$ of the different  
inhomogeneous regions. Consequently, we may use the GL theory to calculate the upper critical field of these 
different superconducting regions which form the whole condensate, and the total sample $H_{c2}$ is the 
sum of these individual inhomogeneous contributions as in Eq.(\ref{4}).

\section{Comparison with experimental data}

In this section  we compare
the results of Eq.(\ref{4}) with the experimental upper critical field 
data of Ref.~\cite{Positivecurve4} and Ref.~\cite{Yayu} of the LSCO family and near optimum Bi2212.

\begin{figure}[h]
\begin{center}
\includegraphics[height=6.5cm]{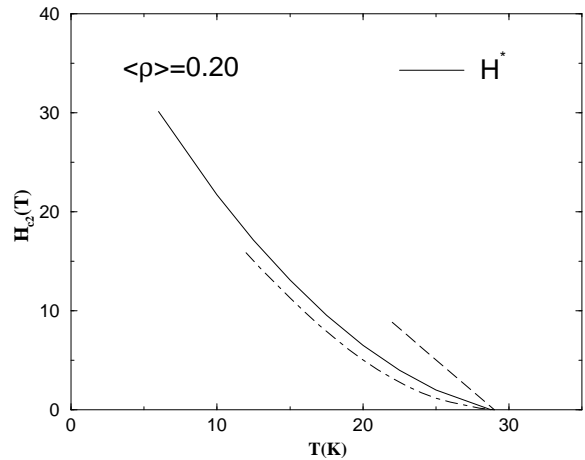}
\caption{Theoretical results of $H_{c2}(T)$ (dot-dashed line) 
for the $\langle \rho \rangle=0.20$ of the LSCO series
together with the Nersnt signal measurements curve of 
Ref.~\cite{Yayu} (solid line). The dashed line is a GL fitting.}
\label{Fig4}
\end{center}
\end{figure} 

The experimental upper critical field $H_{c2}$ of the HTSC may be obtained
from the resistivity measurements as it is the field relative 
to a fraction of the ``normal-state'' resistivity~\cite{Pav,Positivecurve4}. 
The correct fraction which leads to the upper critical field is still controversial and 
there is no conclusion in whether $H_{c2}$ may correspond to the beginning, 
the middle, or the top (end) of the resistivity curves~\cite{Positivecurve4,Positivecurve2}. 
Another way to obtain $H_{c2}$ is from the field dependence of the transport line-entropy 
derived from the Nernst signal~\cite{Yayu}.
Despite of this difficulty we attempt here to compare the theoretical results 
with the available experimental data.
For this purpose we identified some applied magnetic fields of Ref.~\cite{Positivecurve4} 
and $H^*$ of Ref.~\cite{Yayu} as the upper critical fields for 
the selected compounds studied due to the reasons below. 

\begin{figure}[h]
\begin{center}
\includegraphics[height=6.5cm]{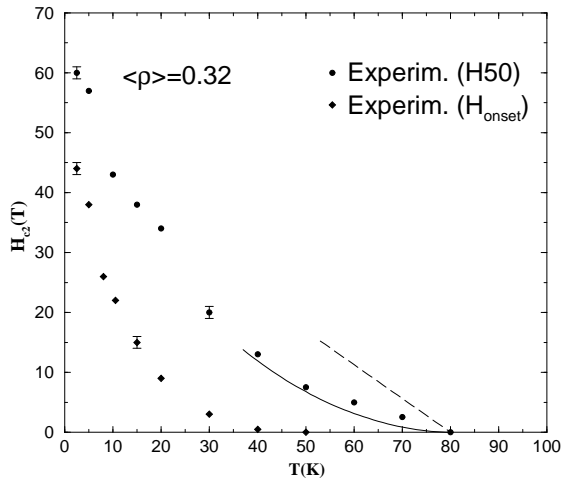}
\caption{Theoretical results of $H_{c2}(T)$ (solid line) 
of the near optimum Bi2212
together with the experimental points of
Ref.~\cite{Positivecurve4}. A GL fitting using Eq.(\ref{3}) is also shown (dashed line) for 
comparison.}
\label{Fig3}
\end{center}
\end{figure}

By definition, $H_{onset}$ from the resistive measurements of Ref.~\cite{Positivecurve4} 
is defined as the magnetic field at which
the resistivity $\rho$ first is detected to deviate from the zero in the $\rho$ vs $H$ curves, 
and this is the assumption used in Eq.(\ref{4}) and, therefore, it is 
our definition of $H_{c2}(T)$. Furthermore, it is reasonable to  
take $T_c$ from the $H$-$T$ phase diagram as the temperatures where $H_{onset}$=0.  
However, it is experimentally observed~\cite{Positivecurve4} that for some compounds $H_{onset}$ vanishes at 
temperatures much smaller than the known  
values of $T_c$. This is because the superconducting transition is sharp at low fields, but spreads itself 
over a large interval of temperatures as the field increases. As a consequence one defines the field $Hx$, 
$x$ being a percentage of the normal state resistivity. In some cases the difference between $H_{onset}$ and 
$H90$ ($90\%$ of the normal-sate resistivity) may be about $50\%$ of $T_c$~\cite{Positivecurve4}. 
When this is the case the correct $H_{c2}$ may be between these two fields. In our calculations 
from Eq(\ref{4})
$H_{c2}$ is calculated for $T\le T_c$ and always vanishes at $T_c$, with $H_{c2}$ being in principle 
the $H_{onset}$ field. 
By the same token, for the Nernst signal measurements of Ref.~\cite{Yayu}, $H^*$ may be considered 
the upper critical field since it represents 
an intrinsic field which controls the onset of the flux-flow dissipation
and vanishes at a temperature close to $T_c$. Therefore, 
$H^*$ may be compared with $H_{onset}$. In agreement with Wang {\it et al.}~\cite{Yayu}, 
attempts to find $H_{c2}$ using the resistivity invariably turn up a curve akin to $H^*$.

Now, in order to compare with the experimental fields of Ref.~\cite{Positivecurve4}, we 
plot $H_{c2}(T)$ with the measured $H_{onset}$ for the cases of 
$\langle \rho \rangle $=0.15 (Fig. \ref{Fig2}a) and 
$\langle \rho \rangle$=0.17 (Fig. \ref{Fig2}b) of the LSCO series. 
Also, in Fig. \ref{Fig2}a we plot the results of a constant and a linear charge distribution 
together with the bimodal distribution of Ref.~\cite{Evandro1}.
As one can see, the distributions yield very similar results, which shows that the calculations 
do not depend on the details of the charge distribution, although without a distribution of 
$T_c(i)$'s we simply obtain a GL linear behavior.

For the case of $\langle \rho \rangle $=0.08 
(Fig. \ref{Fig2}c) we compared our results with $H90$ since this field vanishes at $T_c\approx 24K$, 
which is the value of $T_c$ obtained from the phase diagram of Ref~\cite{Evandro1}, 
while $H_{onset}$ vanishes at $T\approx 12K$. 
In Fig. \ref{Fig4} one can see the results for $\langle \rho \rangle$=0.20 compared with $H^*$
from the Nernst-signal measurements of Ref.~\cite{Yayu}.
For the near-optimum Bi2212 we compared our results 
with $H50$ (Fig. \ref{Fig3}) of Ref.~\cite{Positivecurve4},
which vanishes at $T_c\approx 80K$ and is in accordance 
with the phase diagram of Ref.~\cite{Evandro1}, while 
$H_{onset}$ vanishes at $T\approx50K$. Also, in Fig. \ref{Fig2}c and Fig. \ref{Fig3} the experimental 
points of $H_{onset}$ are shown for comparison.
For the LSCO series a coherence length of $\xi_{ab}(0)$=$30\AA$ was adopted, which is in accordance with
the measurements of  Refs.~\cite{Pav,Qiang,Cooper}.
This value of $\xi_{ab}(0)$ leads to $H_{c2}(0)$=$\Phi/2\pi\xi_{ab}^2(0)$=32T.
For the Bi2212 a coherence length of $\xi_{ab}(0)$=$27\AA$ was considered, which is in accordance with 
Ref.~\cite{Cooper}. Similarly one gets $H_{c2}(0)$=45T. These discrepancies for the low temperature 
values of $H_{c2}(T)$ is clearly due to the GL expressions (Eqs.(\ref{1})-(\ref{4})), which should not be 
used far from $T_c$. This is the reason why we stop our calculations at temperatures below $T_c$/3. 
At very low temperatures we do not know how to estimate the contributions of the islands with 
$T_c(i)$$\approx$$T_c$. 



\section{Conclusion}

We have calculated the upper critical field $H_{c2}(T)$ for a disordered superconductor characterized 
by a distribution of different local critical temperatures $T_c(i)$ at different domains. 
We have applied a simple GL expression to each of these superconducting regions. 
With this procedure we can explain the magnetic signals below and 
above $T_c$. We have been able to fit the $H_{c2}$ curves derived by two different experimental 
procedures, namely the resistive magnetic fields and the Nernst signal~\cite{Yayu1}. In all the 
cases the curves exhibit a positive curvature which is different from the magnetically determined 
$H_{c2}$ lines~\cite{Qiang} from pure GL (Eq.(\ref{3})).
This positive curvature reflects the GL behavior of each individual domain in a disordered superconductor. 
Furthermore, taking the inhomogenous charge distribution into account several properties like 
the Meissner and Nernst effects, which are seen at temperatures much higher than $T_c$,
are naturally explained. Such inhomogeneities are taken into account by a charge distribution, 
but as discussed above, the main features of our calculations are independent of the details of the 
probability charge distribution. 
In conclusion it is crucial to take into consideration the fact that the HTSC are 
inhomogeneous materials in order to describe the main qualitative features of the high-$T_c$ superconductors.



\begin{references}



\bibitem{Bednorz}
J.G. Bednorz and K.A. M\"uller, 
Z. Phys. {\bf 64} 189 (1986).




\bibitem{Positivecurve4}
Y. Ando, G.S. Boebinger, A. Passner, L.F. Schneemeyer, T. Kimura, M. Okuya, S. Watauchi,
J. Shimoyama, K. Kishio, K. Tamasaku, N. Ichikawa, and S. Uchida,
Phys. Rev. B {\bf 60}, 12475 (1999).

\bibitem{Positivecurve2}
M.S. Osofsky, R.J. Soulen, Jr., S.A. Wolf, J.M. Broto, H. Rakoto, J.C. Ousset,
G. Coffe, S. Askenazy, P. Pari, I. Bozovic, J.N. Eckstein, and G.F. Virshup,
Phys. Rev. Lett. {\bf 71}, 2315 (1993).

\bibitem{Positivecurve1}
A.P. Mackenzie, S.R. Julian, G.G. Lonzarich, A. Carrington, S.D. Hughes, R.S. Liu,
D.C. Sinclair, Phys. Rev. Lett. {\bf 71}, 1238 (1993).

\bibitem{Positivecurve3}
G. Kotliar, C.M. Varma, Phys. Rev. Lett. {\bf 77}, 2296 (1996).


\bibitem{Blatter}
G. Blatter, M.V. Feigel'man, V.B. Geshkenbein, A.I. Larkin. and V.M. Vinokur,
Rev. Mod. Phys. {\bf 66}, 1125 (1994).

\bibitem{WHH1}
E. Helfand and N.R. Werthamer,
Phys. Rev. Lett. {\bf 13}, 686 (1964).

\bibitem{WHH2}
N.R. Werthamer, E. Helfand, and P.C. Hohenberg,
Phys. Rev. {\bf 147}, 295 (1966).

\bibitem{Kotliar}
G. Kotliar and A. Kapitulnik,
Phys. Rev. B {\bf 33}, 3146 (1986).

\bibitem{Ovchinnikov}
Yu. N. Ovchinnikov and V. Z. Kresin,
Phys. Rev. B {\bf 52}, 3075 (1995).

\bibitem{Spivak}
B. Spivak and F. Zhou,
Phys. Rev. Lett. {\bf 74}, 2800 (1995).

\bibitem{Geshkenbein}
V. B. Geshkenbein, L. B. Ioffe, and A. J. Millis, 
Phys. Rev. Lett. {\bf 80}, 5778 (1998).

\bibitem{Domanski}
T. Domanski, M. M. Ma\'ska, and M. Mierzejewski 
Phys. Rev. B {\bf 67}, 134507 (2003).

\bibitem{Schofield}
A. J. Schofield,
Phys. Rev. B {\bf 51}, 11733 (1995).

\bibitem{Wheatley}
R. G. Dias, and J. M.  Wheatley,
Phys. Rev. B {\bf 50}, 13887 (1994).

\bibitem{Alexandrov}
A. S. Alexandrov,
Phys. Rev. B {\bf 48}, 10571 (1993).

\bibitem{Maska}
M. Mierzejewski and M. M. Ma\'ska,
Phys. Rev. {\bf B 66}, 214527 (2002). 

\bibitem{Pan}
S.H. Pan, J.P. O'Neal, R.L. Badzey, C. Chamon, H. Ding, J.R. Engelbrecht, Z. Wang,
H. Eisaki, S. Uchida, A.K. Gupta, K.W. Hudson, K.M. Lang, J.C. Davis, Nature {\bf 413},
282-285 (2001), and cond-mat/0107347.

\bibitem{Lang}
K.M. Lang, V. Madhavan, J.E. Hoffman, E.W. Hudson, H. Eisaki, S. Uchida, and J.C. Davis,
Nature {\bf 415}, 412 (2002).

\bibitem{Howald}
C. Howald, P. Fournier, and A. Kapitulnik, 
Phys. Rev. {\bf B 64}, 100504 (2001).

\bibitem{Egami2}
T. Egami and S.J.L. Billinge, in {\it Physical Properties of High Temperature Superconductors V},
edited by D.M. Ginzberg (World Scientific, Singapore, 1996) p.265.

\bibitem{Billinge}
S.J.L. Billinge, Proceedings of the Conference on Major Trends in Superconductivity in 
New Milenium [J. Supercond. (2000)].

\bibitem{Bozin}
E.S. Bozin, G.H. Kwei, H. Takagi, and S.J.L. Billinge, 
Phys. Rev. Lett. {\bf B 84}, 5856 (2000).
 
\bibitem{Egami}
T. Egami, Proc. of the New3SC International Conference, 
Physica C {\bf 364-365}, 561 (2001),, and references there in.

\bibitem{Evandro2}
E.V.L. de Mello, M.T.D. Orlando, J.L. Gonz\'alez, E.S. Caixeiro, E. Baggio-Saitovich,
Phys. Rev. {\bf B 66}, 092504 (2002).

\bibitem{Evandro1}
E.V.L. de Mello, E.S. Caixeiro, and J.L. Gonz\'alez,
Phys. Rev. {\bf B 67}, 024502 (2003).

\bibitem{Timusk}
T. Timusk, and B. Statt, 
Rep. Prog. Phys. {\bf 62}, 61 (1999).

\bibitem{Iguchi}
I. Iguchi, I. Yamaguchi, and A. Sugimoto, 
Nature (London) {\bf 412}, 420 (2001).

\bibitem{Lascialfari}
A. Lascialfari, A. Rigamonti, L. Romano, P. Tedesco, A. Varlamov, and D. Embriaco,
Phys. Rev. {\bf B 65}, 144523 (2002).

\bibitem{Yayu1}
Y. Wang, Z.A. Xu, T. Kakeshita, S. Uchida, S. Ono, Y. Ando, and N.P. Ong,
Phys. Rev. {\bf B 64}, 224519 (2001).

\bibitem{Yayu}
Y. Wang, N.P. Ong, Z.A. Xu, T. Kakeshita, S. Uchida, D.A. Bonn, and W.N. Hardy,
Phys. Rev. Lett {\bf  88}, 257003 (2002) .

\bibitem{Trancada}
J.M. Trancada, B.J. Sternlieb, J.D. Axe, Y. Nakamura and S. Uchida, 
Nature (London) {\bf 375}, 561 (1995).

\bibitem{Bianconi}
A. Bianconi, N.L. Saini, A. Lanzara, M. Missori, T. Rossetti, H. Oyanagi, H. Yamaguchi, K. Oka
and T. Ito, 
Phys. Rev. Lett. {\bf 76}, 3412 (1996).

\bibitem{Xu}
Z.A. Xu, N.P. Ong, Y. Wang, T. Kakeshita, and S. Ushida, 
Nature  {\bf 406}, 486(2000).

\bibitem{Oda}
 Oda, N. Momono, and M. Ido, 
Supercond. Sci. Technol. {\bf 13}, R.139 (2000) 

\bibitem{Fisher}
D.S. Fisher, M.P.A. Fisher, and D.A. Huse, 
Phys. Rev. B {\bf 43}, 130 (1991).

\bibitem{Comprimcorr}
L.N. BulaevskII,
Int. J. Mod. Phys. B {\bf 4}, 1849 (1990).


\bibitem{LD}
W.E. Lawrence, and S. Doniach, in {\it Proc. of the 12th Int. Conf. of Low-Temperature
Physics}, Kyoto, 1970, ed. E. Kanda (Keigaku, 1970), p. 361.

\bibitem{Pav}
M. Cyrot and D. Pavuna, {\it Introduction to Superconductivity and High-$T_c$ Materials}
(World Scientific, 1992)

\bibitem{Franz}
M. Franz, C. Kallin, A.J. Berlinsky, and M.I. Salkola,
Phys. Rev. B {\bf 56}, 7882 (1997).


\bibitem{Qiang}
Q. Li, M. Suenaga, T. Kimura, and K. Kishio,
Phys. Rev. B {\bf 47}, 11384 (1993).

\bibitem{Landau1}
I.L. Landau, and H.R. Ott,
Phys. Rev. {\bf B 66}, 144506 (2002).

\bibitem{Landau2}
I.L. Landau, and H.R. Ott,
Physica C {\bf 385}, 544 (2003).

\bibitem{Ginzburg}
V.L. Ginzburg,
Zh. Eksp. Teor. Fiz. {\bf 23}, 236 (1952).

\bibitem{Uemura}
Y. J. Uemura,
Solid State Commun. {\bf 126}, 23 (2003). 


\bibitem{Cooper}
S.L. Cooper, and K.E. Gray, in {\it Physical Properties of High Temperature Superconductors IV}, 
edited by Donald M. Ginsberg (World Scientific, Singapore), p. 122.



\end{references}
\end{document}